\documentclass{article}

\usepackage{PRIMEarxiv}

\usepackage[utf8]{inputenc} 
\usepackage[T1]{fontenc}    
\usepackage{hyperref}       
\usepackage{url}            
\usepackage{booktabs}       
\usepackage{amsfonts}       
\usepackage{nicefrac}       
\usepackage{microtype}      
\usepackage{lipsum}
\usepackage{fancyhdr}       
\usepackage{graphicx}       
\graphicspath{{media/}}     

\usepackage{amsmath}
\usepackage{stmaryrd}
\usepackage{tikz}
\usepackage{amssymb}
\usetikzlibrary{shapes.geometric, arrows, positioning, calc}

\title{Separation Logic for Verifying Physical Collisions of CNC Programs
}

\author{
  Yeonseok Lee \\
  SLING AI Inc. \\
  Incheon, Republic of Korea \\
  \texttt{ylee@sling.ai.kr}
}

\begin{document}

\newcommand{\bigast}{\scalebox{1.5}{$\ast$}}

\maketitle

\begin{abstract}
Safety verification in Computer Numerical Control (CNC) machining has traditionally relied on simulation-based methods that require repetitive tests when requirements change. This paper introduces a formal verification framework that conceptualizes the physical CNC workspace as a Spatial Heap, treating physical occupancy as a managed logical resource. Central to our approach is a Parser-Prover Handshake that decouples machine kinematics from formal logic. By mapping tool trajectories and safety buffers into a discrete spatial model prior to evaluation, the framework enables the use of Separation Logic (SL) to verify safety via formal triples. Within this model, physical collisions are redefined as logical Spatial Data Races, detected through the failure of the separating conjunction to establish disjointness. Furthermore, we extend the methodology to collaborative environments using Concurrent Separation Logic (CSL), where physical hand-offs are verified as formal ownership transfers. Additionally, the framework scales to multi-axis kinematics (e.g., 5-axis Table-Table configurations) by treating the workpiece as a dynamically mutable spatial resource. By serving as a complement to traditional geometric simulation, this approach reduces the number of required iterative test cycles, offering a foundation for autonomous, less-collision manufacturing.
\end{abstract}

\keywords{CNC Manufacturing \and G-code \and Separation Logic \and Formal Verification \and Collision Avoidance \and Concurrency \and Multi-axis CNC}

\section{Introduction}
The verification of trajectory safety in Computer Numerical Control (CNC) machining represents a critical intersection between computational geometry and formal methods. Traditional safety verification has historically relied upon geometric simulations and swept volume analysis that validate specific paths through a virtual representation of the workspace. While such simulations are effective for catching obvious errors in G-code, they are fundamentally limited by their inability to provide symbolic proofs of safety and often struggle with recursive feasibility in dynamic environments. 
Furthermore, these methods are often computationally expensive, requiring repetitive testing to establish G-code changes whenever specifications change.

This paper provides an analysis of a deterministic verification methodology that leverages Separation Logic (SL) \cite{reynolds2002separation, berdine2005symbolic, o2004resources} to model the physical CNC workspace as a logical memory resource. While recent work explores the mechanized semantics of G-code for additive manufacturing \cite{tekriwal2025mechanized}, our approach focuses specifically on collision avoidance by establishing a formal correspondence between physical occupancy and logical memory ownership. This framework facilitates the detection of collisions as logical contradictions, termed \textit{Spatial Data Races}.

The evolution of program verification has been profoundly shaped by the development of the axiomatic basis for programming \cite{hoare1969axiomatic} and the subsequent emergence of Separation Logic (SL), a Hoare-style framework designed to reason about programs that manipulate shared mutable data structures \cite{reynolds2002separation}. Separation Logic has provided a scalable solution to the ``frame problem''—the challenge of describing what a program component does not change. By introducing the separating conjunction ($*$), which asserts that its subformulas hold for disjoint portions of memory, SL enables ``local reasoning'' \cite{o2004resources}. This means that the specification and proof of a program component mention only the portion of memory accessed by that component, independent of the rest of the global state.

While Separation Logic was originally intended for the domain of software memory safety, its underlying principles of resource separation are uniquely suited to the verification of cyber-physical systems (CPS), specifically CNC machining. In a physical CNC workspace, safety is defined by the requirement that different objects—such as the cutting tool, the workpiece (stock), and the machine fixtures (environment)—never occupy the same physical coordinates simultaneously. Current industry standards for validating these requirements primarily involve geometric simulation and probabilistic modeling. These simulations, however, are often stochastic, providing only a high degree of confidence rather than a deterministic proof. They fail to account rigorously for the continuous kinematic variables that characterize real-world machine behavior, such as non-deterministic rapid-positioning paths and mechanical variability.

This paper introduces a deterministic alternative that conceptualizes the physical CNC workspace as a managed logical resource termed a \textit{Spatial Heap}. By mapping physical occupancy to logical memory ownership, we redefine physical collisions as \textit{Spatial Data Races}—logical contradictions that arise when two distinct physical resources attempt to claim ownership of the same discrete memory address simultaneously. We propose a framework that decouples the continuous kinematics of the machine from the discrete logic of the proof engine through a formal \textit{Parser-Prover Handshake}. This architectural boundary allows the system to exhaust physical uncertainties into a discrete integer domain before evaluation, bypassing the high computational complexity typically associated with general pointer arithmetic in separation logic systems \cite{brotherston2018complexity} and focusing on more tractable fragments of the logic \cite{cook2011tractable}.

Our contributions include the adaptation of symbolic execution techniques from Berdine et al. \cite{berdine2005symbolic} to evaluate spatial footprints and the integration of geometric expansion via Minkowski sums to provide formal safety margins ($\epsilon$) \cite{lozano1983spatial, lien2009simple, minarvcik2024minkowski}. Furthermore, we extend the framework to concurrent machining environments using Concurrent Separation Logic (CSL) \cite{brookes2016concurrent}, enabling the verification of collaborative robotic cells and multi-tool systems. Crucially, our methodology is not limited to foundational 3-axis operations. By conceptualizing the physical stock as an independent, dynamically mutable spatial entity, the framework natively scales to multi-axis kinematics (e.g., 5-axis Table-Table configurations). The tool and the workpiece are modeled as concurrent spatial threads moving through an absolute coordinate system.
By serving as a scalable, mathematically grounded complement to traditional geometric simulation, this approach  minimizes the number of required iterative test cycles, offering a  foundation for autonomous, less-collision manufacturing.

\section{Architectural Framework}

The methodology is founded on a strict architectural decoupling between the evaluation of physical kinematics and the formal proof of spatial disjointness. By establishing a formal boundary between a continuous-domain Parser and a SL Prover, the system ensures that physical safety is evaluated as a deterministic logical property. 
This separation is strategically designed to avoid the high computational complexity associated with general pointer arithmetic in Separation Logic \cite{brotherston2018complexity}, thereby allowing the Prover to operate within a more tractable fragment of the logic \cite{cook2011tractable}.

\subsection{Verification Pipeline}

The transition from raw machine commands to verified safety metadata is modeled as a unidirectional pipeline, building upon recent efforts to establish mechanized semantics for CNC languages \cite{tekriwal2025mechanized}. In this architecture, the Parser acts as a virtual machine that resolves continuous inputs into discrete sets, which are then consumed by the Prover to validate the Spatial Heap.

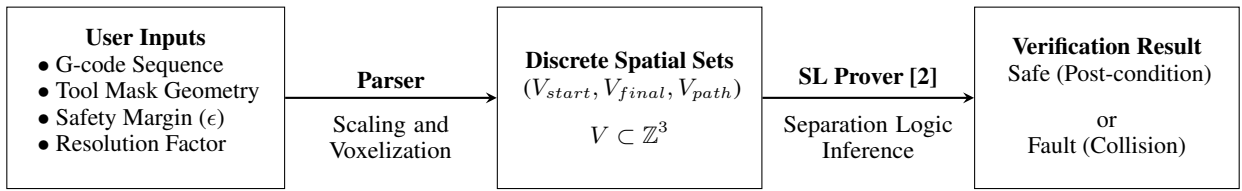
\begin{figure}[h]
    \centering
    \begin{tikzpicture}[
        box/.style={
            rectangle, 
            draw, 
            minimum width=3.5cm, 
            minimum height=2.4cm, 
            align=center, 
            font=\small
        },
        arrow/.style={thick, ->, >=stealth},
        label_text/.style={font=\footnotesize, align=center, text width=3cm}
    ]

    \node (inputs) [box] {
        \textbf{User Inputs} \\
        \vspace{0.1cm}
        \begin{tabular}{l}
        $\bullet$ G-code Sequence \\
        $\bullet$ Tool Mask Geometry \\
        $\bullet$ Safety Margin ($\epsilon$) \\
        $\bullet$ Resolution Factor
        \end{tabular}
    };

    \node (sets) [box, right=2.8cm of inputs] {
        \textbf{Discrete Spatial Sets} \\
        \vspace{0.2cm}
        ($V_{start}, V_{final}, V_{path}$) \\
        $V \subset \mathbb{Z}^3$
    };

    \node (result) [box, right=2.8cm of sets] {
        \textbf{Verification Result} \\
        \vspace{0.2cm}
        Safe (Post-condition) \\
        or \\
        Fault (Collision)
    };

    \draw [arrow] (inputs) -- node[anchor=south, font=\small\bfseries] {Parser} 
                               node[anchor=north, label_text, yshift=-0.1cm] {Scaling and \\ Voxelization} (sets);
    
    \draw [arrow] (sets) -- node[anchor=south, font=\small\bfseries] {SL Prover \cite{berdine2005symbolic}} 
                             node[anchor=north, label_text, yshift=-0.1cm] {Separation Logic \\ Inference} (result);

    \end{tikzpicture}
    \caption{System Architecture Pipeline for CNC Verification.}
\end{figure}

\subsection{The Parser-Prover Handshake}

The architecture enforces a mathematical boundary to prevent the logical domains from becoming complicated by continuous kinematic variables, a necessary step to maintain decidability in spatial reasoning.

\begin{itemize}
    \item \textbf{The Parser:} The Parser handles the pre-compilation phase. It consumes the raw G-code and tool geometry, applying a discretization process to translate floating-point values into unique integer addresses. By the time the Parser emits its output, the logical Store is exhausted, and all variables are replaced by concrete spatial literals.
    \item \textbf{The SL Prover:} The Prover operates as a discrete memory manager. It takes the discrete coordinate sets produced by the Parser and evaluates them against the Spatial Heap. It uses the Separating Conjunction ($*$) to ensure that the tool's claimed memory region is disjoint from the environment fixtures, treating physical space as a memory resource \cite{reynolds2002separation, o2004resources}.
\end{itemize}

\begin{figure}[h]
    \centering
    \begin{tikzpicture}[
        box/.style={
            rectangle, 
            draw, 
            minimum width=3.8cm, 
            minimum height=2.2cm, 
            align=center, 
            font=\small
        },
        heading/.style={
            align=center, 
            font=\small\bfseries
        },
        arrow/.style={thick, ->, >=stealth},
        biarrow/.style={thick, <->, >=stealth},
        long_label/.style={font=\scriptsize, align=center, text width=3.8cm}
    ]

    \node (cont_title) at (0,4) [heading] {
        [ CONTINUOUS DOMAIN ] \\
        (Pre-compilation)
    };
    \node (disc_title) at (7,4) [heading] {
        [ DISCRETE DOMAIN ] \\
        (Logical Inference)
    };

    \node (store) [box, below=0.4cm of cont_title] {
        \textbf{Store ($s$)} \\
        \vspace{0.1cm}
        Maps variables to \\
        kinematics: \\
        $X$, $Y$, $Z$, or $F$
    };

    \node (heap) [box, below=0.4cm of disc_title] {
        \textbf{Spatial Heap ($h$)} \\
        \vspace{0.1cm}
        Maps $\mathbb{Z}^3$ voxels to \\
        physical states: \\
        Tool, Stock, Env, or Empty        
    };

    \node (parser) [box, below=1.8cm of store] {
        \textbf{The Parser}
    };

    \node (prover) [box, below=1.8cm of heap] {
        \textbf{The SL Prover}
    };

    \draw [arrow] (store) -- node[anchor=south, font=\footnotesize\bfseries] {Exhaustion} 
                             node[anchor=north, long_label, yshift=-0.1cm] {All variables are replaced \\ by static spatial literals} (heap);

    \draw [biarrow] (parser) -- node[anchor=west, font=\footnotesize, align=left, xshift=0.1cm] {(Dynamic \\ Updates)} (store);
    \draw [biarrow] (prover) -- node[anchor=east, font=\footnotesize, align=right, xshift=-0.1cm] {(Logical \\ Proof)} (heap);

    \end{tikzpicture}
    \caption{Conceptual Framework: Decoupling Memory Domains.}
\end{figure}
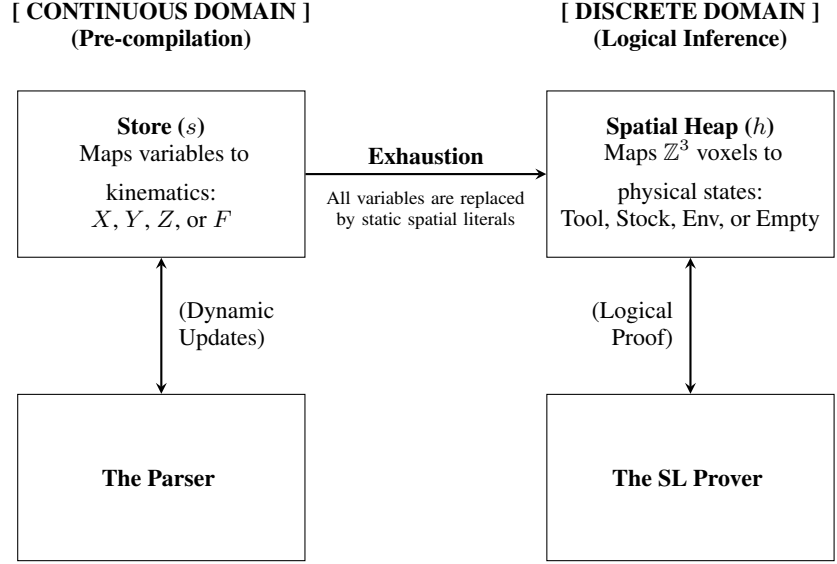

\subsection{Memory Domains: Store vs. Spatial Heap}

The system delineates two distinct memory domains to maintain computational viability, mirroring the standard model used in foundational Separation Logic \cite{reynolds2002separation}:
\begin{enumerate}
    \item \textbf{Store ($s$):} A transient mapping used exclusively by the Parser to track kinematic identifiers and system variables. The Store is fully resolved and emptied before the proof phase begins.
    \item \textbf{Spatial Heap ($h$):} A persistent logical model managed by the Prover. Every coordinate is mapped to a physical occupancy state: Tool, Environment, Stock, or Empty. Ownership of these ``spatial voxels'' is managed as a discrete, finite memory resource.
\end{enumerate}

\section{Modeling Physical Uncertainty}

\subsection{Swept Volume and Geometric Expansion}

To guarantee safe spatial execution, the nominal tool volume is expanded by a worst-case safety margin $\epsilon$. This process leverages the concept of swept volumes \cite{abrams2000computing} and configuration space modeling \cite{lozano1983spatial} to ensure that every potential point of contact is accounted for. Rather than modeling continuous machine-specific kinematics—such as servo lag or acceleration profiles—directly within the logic, $\epsilon \in \mathbb{Z}^+$ is provided as a user-defined discrete parameter. This represents the number of spatial units (voxels) by which the tool's volume must be padded to achieve robust safety guarantees in dynamic environments \cite{stamouli2024recursively}. 

By defining $\epsilon$ directly within the discrete integer domain, we maintain strict type safety between the parser and the Separation Logic prover. To define this geometric dilation, we utilize a Minkowski sum approach \cite{lien2009simple, minarvcik2024minkowski}, defining $B_\epsilon$ as a discrete axis-aligned bounding box using the $L_{\infty}$ Chebyshev norm:

\begin{equation}
B_\epsilon \triangleq \{ c \in \mathbb{Z}^3 \mid \|c\|_\infty \leq \epsilon \} = \{ (x,y,z) \in \mathbb{Z}^3 \mid \max(|x|, |y|, |z|) \leq \epsilon \}
\end{equation}

\subsection{Definition: Path Functions}
The function $\text{path}: \mathbb{Z}^3 \times \mathbb{Z}^3 \rightarrow 2^{\mathbb{Z}^3}$ generates the geometric locus of all intermediate coordinates between a starting point ($c_{start}$) and a target ($c_{target}$). Under the discrete spatial heap model, these paths are evaluated as finite sets of voxel addresses.

\subsection{Definition: Linear Path Function}
The function $path_{lin}: \mathbb{Z}^{3} \times \mathbb{Z}^{3} \rightarrow 2^{\mathbb{Z}^{3}}$ generates a finite discrete trajectory for linear interpolation (\texttt{G01}). This ensures all intermediate coordinates $c$ are strictly evaluated and bounded within the discrete integer domain to maintain consistency with the spatial heap:
\begin{equation}
    path_{lin}(c_{start}, c_{target}) \triangleq \text{Bresenham3D}(c_{start}, c_{target})
\end{equation}

\subsection{Discrete Trajectory Generation: Bresenham 3D}

The translation of linear interpolation commands into the discrete spatial heap relies on an extension of the algorithm originally proposed by Bresenham \cite{bresenham1998algorithm}. While the original algorithm was designed for 2D digital plotters, its 3D extension is utilized here to generate a sequence of voxel coordinates that minimize the error between the discrete path and the ideal continuous line segment.

This algorithm is particularly suited for our framework because it operates entirely within the integer domain ($\mathbb{Z}^3$), avoiding the introduction of floating-point inaccuracies during the "Parser" phase. For a move from $c_{start}$ to $c_{target}$, we define the driving dimensions $\Delta x, \Delta y, \Delta z$ and the corresponding error terms. The function $path_{lin}$ iteratively selects the next voxel $c_{i+1}$ such that:
\begin{equation}
c_{i+1} = c_i + \text{step}(\delta_x, \delta_y, \delta_z)
\end{equation}
where the decision variables $\delta$ are updated via integer addition and subtraction based on the cumulative deviation from the ideal slope. By employing this method, the Parser "exhausts" the continuous trajectory into a finite set of memory addresses $V_{path} = \{ c_0, c_1, \dots, c_n \} \subset \mathbb{Z}^3$.

\subsection{Definition: Box Path Function}
The function $path_{box}: \mathbb{Z}^3 \times \mathbb{Z}^3 \rightarrow 2^{\mathbb{Z}^3}$ generates the worst-case Cartesian bounding box of all intermediate coordinates between a discrete starting point and a target. 
Let the relational operators $\leq$, $\min$, and $\max$ over $\mathbb{Z}^3$ be defined as pointwise operations. 
For any $a, b \in \mathbb{Z}^3$ where $a = (a_x, a_y, a_z)$ and $b = (b_x, b_y, b_z)$:

\begin{align*}
    a \leq b  &\iff  a_x \leq b_x \wedge a_y \leq b_y \wedge a_z \leq b_z \\[10pt]
    \min( a , b  )  &=  
    (\min(a_x, b_x), \min(a_y, b_y), \min(a_z, b_z)) \\[10pt]
    \max( a , b  )  &=  
    (\max(a_x, b_x), \max(a_y, b_y), \max(a_z, b_z))
\end{align*}

This formulation accounts for uncoordinated, machine-dependent motion during rapid multi-axis travel (\texttt{G00}), where the exact trajectory is non-deterministic and can be conservatively approximated by its bounding volume:
\begin{equation}
path_{box}(c_{start}, c_{target}) \triangleq \{ c \in \mathbb{Z}^3 \mid \min(c_{start}, c_{target}) \leq c \leq \max(c_{start}, c_{target}) \}
\end{equation}

\section{The Parser: Static Evaluation and Discretization}

The architecture enforces a strict boundary between the \textbf{Parser} (Virtual Machine) and the \textbf{SL Prover} (Memory Manager). The Parser is responsible for exhausting continuous kinematic variables and physical uncertainties, emitting a purely discrete set of spatial assertions and command triples for the Prover to evaluate.

\subsection{Raw G-Code Input Syntax}
The Parser ingests imperative instructions from industrial CNC programs. The raw input language consists of parameter assignments and standard motion codes:
\begin{itemize}
    \item \textbf{Assignment ($\texttt{X} = n$):} Updates a continuous kinematic parameter in the Store ($s$).
    \item \textbf{Rapid Positioning (\texttt{G00 X Y Z}):} Commands non-cutting motion to target coordinates.
    \item \textbf{Linear Interpolation (\texttt{G01 X Y Z F}):} Commands a linear cutting motion at a specific feed rate ($F$).
\end{itemize}

\subsection{Multiplier Discretization}

The logical Store ($s$) is formally defined as a partial function mapping continuous kinematic variables (such as coordinate axes or feed rates) and system identifiers to discrete values:
\begin{equation}
s : \text{Var} \rightarrow \mathbb{Z} \cup \mathbb{Z}^3
\end{equation}
By exhausting these continuous variables during the pre-compilation phase, the Store resolves all identifiers into discrete integers. 

To convert floating-point data into the spatial heap's grid, the Parser applies a simplified discretization function $\mathcal{S}_{\text{grid}}$. Given a user-defined scaling multiplier $\mu \in \mathbb{Z}^{+}$, coordinates are scaled and truncated to the integer domain:
\begin{align*}
\mathcal{S}_{\text{grid}}(x) & \triangleq (\lfloor x \cdot \mu \rfloor) 
\\
\mathcal{S}_{\text{grid}}(x,y,z) & \triangleq (\lfloor x \cdot \mu \rfloor, \lfloor y \cdot \mu \rfloor, \lfloor z \cdot \mu \rfloor) 
\end{align*}
This mapping ensures that raw G-code parameters are scaled into a strictly bounded discrete voxel grid, where every unit corresponds to the machine's minimum physical resolution.

\subsection{Geometric Modeling via Minkowski Sums}

The Parser determines the tool's physical footprint by calculating the swept volume $V_{\text{swept}} \subseteq \mathbb{Z}^3 $. 
This is generated using the Minkowski sum ($\oplus$) of the discrete path $path \subseteq \mathbb{Z}^3$ and the nominal tool volume $V_{\text{tool}} \subseteq \mathbb{Z}^3$. 

Formally, for any two finite sets of discrete spatial coordinates $A, B \subseteq \mathbb{Z}^3$, the Minkowski sum $A \oplus B$ is defined as the set resulting from the pointwise vector addition of every element in $A$ with every element in $B$:
\begin{equation}
A \oplus B \triangleq \{ a + b \mid a \in A, b \in B \}
\end{equation}
where the addition of coordinates is evaluated as $a + b = (a_x + b_x, a_y + b_y, a_z + b_z)$. 

Applied to the tool's kinematics, the discrete swept volume traversed by the tool is strictly evaluated as:
\begin{equation}
V_{\text{swept}} = path \oplus V_{\text{tool}} = \{ p + v \mid p \in path, v \in V_{\text{tool}} \}
\end{equation}

By resolving this tool volume expansion statically at the Parser level, the geometric complexities of physical interference are fundamentally abstracted away. Consequently, the SL Prover receives exact, pre-evaluated spatial footprints, entirely removing the need for dynamic geometric calculations during the logical proof phase.

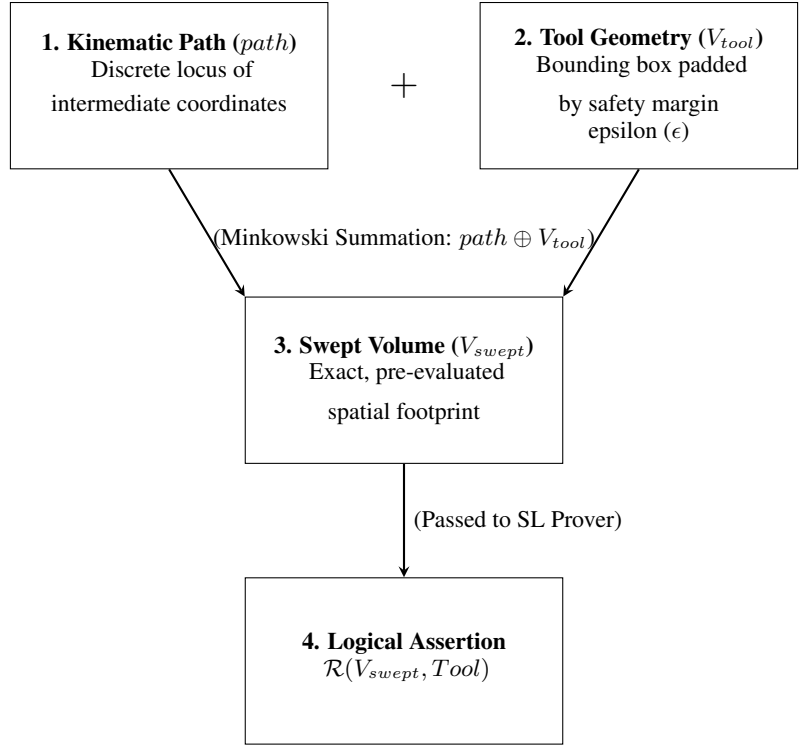
\begin{figure}[h]
    \centering
    \begin{tikzpicture}[
        box/.style={
            rectangle, 
            draw, 
            minimum width=4.2cm, 
            minimum height=2.2cm, 
            align=center, 
            font=\small
        },
        arrow/.style={thick, ->, >=stealth}
    ]

    \node (path) [box] {
        \textbf{1. Kinematic Path ($path$)} \\
        \vspace{0.1cm}
        Discrete locus of \\
        intermediate coordinates \\
    };

    \node (tool) [box, right=2cm of path] {
        \textbf{2. Tool Geometry ($V_{tool}$)} \\
        \vspace{0.1cm}
        Bounding box padded \\
        by safety margin \\
        epsilon ($\epsilon$)
    };

    \node (plus) at ($(path)!0.5!(tool)$) {\Large $\mathbf{+}$};

    \node (swept) [box, below=2.5cm of plus] {
        \textbf{3. Swept Volume ($V_{swept}$)} \\
        \vspace{0.1cm}
        Exact, pre-evaluated \\
        spatial footprint
    };

    \node (assertion) [box, below=1.5cm of swept] {
        \textbf{4. Logical Assertion} \\
        \vspace{0.1cm}
        $\mathcal{R}(V_{swept}, Tool)$
    };

    \draw [arrow] (path.south) -- (swept.north west);
    \draw [arrow] (tool.south) -- (swept.north east);
    
    \node [above=0.5cm of swept, font=\footnotesize] {
    (Minkowski Summation: $path \oplus V_{tool}$)};

    \draw [arrow] (swept.south) -- node[anchor=west, font=\footnotesize] {(Passed to SL Prover)} (assertion.north);

    \end{tikzpicture}
    \caption{Geometric Modeling: Tool volume expansion via Minkowski Sums.}
\end{figure}

\section{Algorithm for Sequential Compilation}

To transition state-dependent G-code into independent Separation Logic assertions, the Parser operates in two distinct phases: establishing the initial memory layout of the workspace and recursively compiling the continuous kinematics into a discrete sequence. The ultimate output is a  Separation Logic Triple \cite{berdine2005symbolic,cook2011tractable} for the entire program, structured as $\{\Sigma_0\} \ \mathbb{C}_{SL} \ \{\Sigma_{final}\}$, ready for evaluation by the SL Prover.

\subsection{Main Compilation and Spatial Initialization}
Before parsing motion commands, the Parser constructs the initial spatial state ($\Sigma_0$) based on the user's predefined setup. After compiling the command sequence, it defines the expected final spatial state ($\Sigma_{final}$) to complete the formal triple.

\textbf{Algorithm: \texttt{Compile\_Program}($Env_{raw}, Stock_{raw}, Tool_{raw}, \mathbb{C}_{raw}$)}
\begin{enumerate}
    \item \textbf{Discretization of Physical Assets:}
    \begin{itemize}
        \item Let $C_{env} = \mathcal{S}_{grid}(Env_{raw})$
        \item Let $C_{stock} = \mathcal{S}_{grid}(Stock_{raw})$
        \item Let $V_{tool} = \mathcal{S}_{grid}(Tool_{raw})$
    \end{itemize}
    
    \item \textbf{Store Initialization and Tool Placement:}
    \begin{itemize}
        \item Initialize the Store with the machine's home position: $s(pos) = (0,0,0)$
        \item Calculate the initial tool volume footprint: $V_{start} = \{s(pos)\} \oplus V_{tool}$
    \end{itemize}
    
    \item \textbf{Evaluate Empty Space:}
    \begin{itemize}
        \item Define the remaining free space in the bounded workspace $W$: \\
        $C_{empty} = W \setminus (C_{env} \cup C_{stock} \cup V_{start})$
    \end{itemize}

    \item \textbf{Construct Initial Precondition ($\Sigma_0$):}
    \begin{itemize}
        \item The Parser emits the starting memory state using the Iterated Separating Conjunction:
        $$ \Sigma_0 \triangleq \mathcal{R}(V_{start}, \text{Tool}) \ast \mathcal{R}(C_{env}, \text{Environment}) \ast \mathcal{R}(C_{stock}, \text{Stock}) \ast \mathcal{R}(C_{empty}, \text{Empty}) $$
    \end{itemize}

    \item \textbf{Compile Sequence:}
    \begin{itemize}
        \item Let $(\mathbb{C}_{SL}, pos_{final}) = \mathcal{C}_{seq}(\mathbb{C}_{raw}, s(pos), V_{tool})$
    \end{itemize}

    \item \textbf{Construct Final Postcondition ($\Sigma_{final}$):}
    \begin{itemize}
        \item Let $V_{end} = \{pos_{final}\} \oplus V_{tool}$
        \item The postcondition asserts the tool's final location alongside the unconsumed resources (using $\mathbf{true}$ as a spatial wildcard for the remaining environment and mutated stock):
        $$ \Sigma_{final} \triangleq \mathcal{R}(V_{final}, \text{Tool}) \ast \mathbf{true} $$
    \end{itemize}

    \item \textbf{Return the SL Triple:}
    \begin{itemize}
        \item \textbf{Return } $\{ \Sigma_0 \} \ \mathbb{C}_{SL} \ \{ \Sigma_{final} \}$
    \end{itemize}
\end{enumerate}

\subsection{The Recursive Compilation Algorithm}

Once the initial spatial heap is established, the Parser pre-compiles the continuous motion sequence into a static discrete sequence. This algorithm tracks tool kinematics in the Store ($s$) and emits the semicolon-separated sequence of SL-style commands alongside the embedded safety assertions, returning both the command string and the final tool position.

\textbf{Algorithm: $\mathcal{C}_{seq}(\mathbb{C}_{raw}, pos, V_{tool})$}
\begin{enumerate}
    \item \textbf{Base Case:}
    $\mathcal{C}_{seq}(\emptyset, pos, V_{tool}) = (\emptyset, pos)$

    \item \textbf{Variable Assignment (\texttt{X} $= n :: \mathbb{C}_{tail}$):}
    \begin{align*}
        &s(\texttt{X}) \leftarrow n \\
        &\text{Return } \mathcal{C}_{seq}(\mathbb{C}_{tail}, pos, V_{tool})
    \end{align*}

    \item \textbf{Rapid Positioning (\texttt{G00 X Y Z} $:: \mathbb{C}_{tail}$):}
    \begin{itemize}
        \item Let $c_{start} = pos$ and $c_{final} = \mathcal{S}_{grid}(\texttt{X}, \texttt{Y}, \texttt{Z})$
        \item Let $Path_{traj} = path_{box}(c_{start}, c_{final})$
        \item Let $V_{start} = \{c_{start}\} \oplus V_{tool}$ and $V_{final} = \{c_{final}\} \oplus V_{tool}$
        \item Let $V_{path} = Path_{traj} \oplus V_{tool}$
        \item Let $\mathbb{C}_{current} =$ ``$\mathbf{assert}\ \mathcal{R}(V_{start}, \text{Tool}) \ast \mathcal{R}(V_{path} \setminus V_{start}, \text{Empty}) ; G00(V_{start}, V_{final}, V_{path})$''
        \item Let $(\mathbb{C}_{tail}, pos_{end}) = \mathcal{C}_{seq}(\mathbb{C}_{tail}, c_{final}, V_{tool})$
        \item \textbf{Return } $(\mathbb{C}_{current} \ ; \ \mathbb{C}_{tail}, \ pos_{end})$
    \end{itemize}

    \item \textbf{Linear Interpolation 
    (\texttt{G01 X$x$ Y$y$ Z$z$ F$f$} $:: \mathbb{C}_{tail}$):}
    \begin{itemize}
        \item Let $c_{start} = pos$ and $c_{final} = \mathcal{S}_{grid}(x, y, z)$
        \item Let $Path_{traj} = path_{lin}(c_{start}, c_{final})$
        \item Let $V_{start} = \{c_{start}\} \oplus V_{tool}$ and $V_{final} = \{c_{final}\} \oplus V_{tool}$
        \item Let $V_{path} = Path_{traj} \oplus V_{tool}$
        \item Let $V_{cut} = (V_{path} \setminus V_{start}) \cap C_{stock}$
        \item Let $\mathbb{C}_{current} =$ ``$\mathbf{assert}\ \mathcal{R}(V_{start}, \text{Tool}) \ast \mathcal{R}(C_{stock}, \text{Stock}) \ast \mathcal{R}(C_{env}, \text{Environment}) ; \mathbf{assert}\ V_{path} \setminus V_{start} \subseteq C_{stock} \cup C_{empty} ; G01(V_{start}, V_{final}, V_{path}) ; \mathbf{foreach}\ c \in V_{cut}\ \mathbf{do}\ [c] := \text{Empty}$''
        \item Update Parser State: $C_{stock} \leftarrow C_{stock} \setminus V_{cut}$ \text{ and } $C_{empty} \leftarrow C_{empty} \cup V_{cut}$
        \item Let $(\mathbb{C}_{tail}, pos_{end}) = \mathcal{C}_{seq}(\mathbb{C}_{tail}, c_{final}, V_{tool})$
        \item \textbf{Return } $(\mathbb{C}_{current} \ ; \ \mathbb{C}_{tail}, \ pos_{end})$
    \end{itemize}
\end{enumerate}

\subsection{Trace Example: 1D Machining Scenario}

To illustrate the deterministic output of the \texttt{Compile\_Program} algorithm, consider a simplified 1D workspace $W = \{0, 1, 2, 3, 4\}$ with a multiplier $\mu = 1$ and a point tool $V_{tool} = \{0\}$. The input raw G-code sequence ($\mathbb{C}_{raw}$) is:
\begin{verbatim}
    G00 X1
    G01 X3 F100
\end{verbatim}

\subsubsection{Phase 1: Spatial Initialization}
The Parser first evaluates the physical environment to establish the initial precondition $\Sigma_0$. With stock defined at $C_{stock} = \{2, 3\}$ and the tool starting at $pos = 0$, the remaining space is calculated as $C_{empty} = \{1, 4\}$.
\begin{equation*}
    \Sigma_0 \triangleq \mathcal{R}(\{0\}, \text{Tool}) \ast \mathcal{R}(\{2, 3\}, \text{Stock}) \ast \mathcal{R}(\{1, 4\}, \text{Empty})
\end{equation*}

\subsubsection{Phase 2: Recursive Sequence Compilation}
The function $\mathcal{C}_{seq}$ processes the commands sequentially, tracking the state updates in the Store:

\begin{enumerate}
    \item \textbf{Processing \texttt{G00 X1}:}
    \begin{itemize}
        \item $c_{start} = 0, c_{final} = 1 \implies Path_{traj} = \{0, 1\}$.
        \item $V_{start} = \{0\}, V_{final} = \{1\}, V_{path} = \{0, 1\}$.
        \item \textbf{Generated Assertion:} $\mathbf{assert}\ \mathcal{R}(\{0\}, \text{Tool}) \ast \mathcal{R}(\{1\}, \text{Empty})$
        \item \textbf{Command:} $G00(\{0\}, \{1\}, \{0, 1\})$
    \end{itemize}

    \item \textbf{Processing \texttt{G01 X3}:}
    \begin{itemize}
        \item $c_{start} = 1, c_{final} = 3 \implies Path_{traj} = \{1, 2, 3\}$.
        \item $V_{start} = \{1\}, V_{final} = \{3\}, V_{path} = \{1, 2, 3\}$.
        \item \textbf{Generated Assertion 1:} $\mathbf{assert}\ \mathcal{R}(\{1\}, \text{Tool}) \ast \mathcal{R}(\{2, 3\}, \text{Stock})$
        \item \textbf{Generated Assertion 2:} $\mathbf{assert}\ \{2, 3\} \subseteq C_{stock} \cup C_{empty}$
        \item \textbf{Command:} $G01(\{1\}, \{3\}, \{1, 2, 3\})$
    \end{itemize}
\end{enumerate}

\subsubsection{Phase 3: Final Triple Generation}
The Parser concludes by identifying the final position $pos_{final} = 3$ and constructing the complete SL Triple:
\begin{gather*}
    \{ \Sigma_0 \} \\
    \mathbf{assert}\ \dots ; G00(\dots) ; \\
    \mathbf{assert}\ \dots ; \mathbf{assert}\ \dots ; G01(\dots) \\
    \{ \mathcal{R}(\{3\}, \text{Tool}) \ast \mathbf{true} \}
\end{gather*}

\section{The Prover: Spatial Heap and Formal Semantics}
\label{sec:prover}

The methodology is founded on a strict architectural decoupling between the evaluation of physical kinematics and the formal proof of spatial disjointness. The SL Prover operates strictly as a discrete memory manager. By the time the Parser emits its output, the logical Store is exhausted, and all variables are replaced by concrete spatial literals. Consequently, the Prover evaluates these discrete coordinate sets solely against the Spatial Heap ($h$) without dynamically querying a variable store.

\subsection{The Spatial Heap Function and Satisfaction Relations}
\label{subsec:spatial_heap}

To rigorously ground the operational semantics of the Zero-Store Model, it is necessary to formally define the mathematical structure of the spatial heap and the satisfaction relations that govern it. Because the Parser statically exhausts the continuous kinematic variables into discrete sets, the Prover's state is entirely defined by spatial occupancy without dynamically querying a variable store.

\subsubsection{The Spatial Heap Function}

Let the domain of discrete physical locations be $\mathbb{Z}^3$. Let the domain of valid physical object states be defined as $\text{status} = \{\text{Tool, Environment, Stock, Empty}\}$. 

The spatial heap $h$ is modeled as a finite partial function mapping discrete physical coordinates to their specific occupancy status:
\begin{equation*}
    h : \mathbb{Z}^3 \rightharpoonup \text{status}
\end{equation*}

Two spatial heaps, $h_1$ and $h_2$, are considered mathematically disjoint, denoted as $h_1 \perp h_2$, if and only if their defined domains share no common coordinate addresses in $\mathbb{Z}^3$:
\begin{equation*}
    h_1 \perp h_2 \iff \text{dom}(h_1) \cap \text{dom}(h_2) = \emptyset
\end{equation*}

When $h_1 \perp h_2$ holds, their disjoint union is denoted as $h_1 \uplus h_2$. This fundamental property serves as the mathematical bedrock for proving the absence of physical collisions in the workspace, directly supporting the deterministic mutual exclusion mechanism.

\subsubsection{Satisfaction Relations}

The formal satisfaction relation $h \models P$ dictates the precise conditions under which a given spatial heap $h$ satisfies a Separation Logic formula $P$. Adapted for CNC verification, these relations are strictly evaluated over discrete spatial sets rather than traditional memory pointers:

\begin{itemize}
    \item \textbf{Empty Heap}: The heap asserts no ownership over any spatial coordinates.
    \begin{equation*}
        h \models \text{emp} \iff \text{dom}(h) = \emptyset
    \end{equation*}

    \item \textbf{Single Spatial Voxel (Points-to)}: The heap exactly comprises a single discrete coordinate $c$ mapped to the physical state $\theta$.
    \begin{equation*}
        h \models c \mapsto \theta \iff \text{dom}(h) = \{c\} \land h(c) = \theta
    \end{equation*}

    \item \textbf{Spatial Resource Allocation}: The heap comprises the finite coordinate set $C$, where every discrete coordinate $c \in C$ is mapped to the state $\theta$. This is logically equivalent to the iterated separating conjunction ($\bigast$) of individual voxels over the set $C$.
    \begin{equation*}
        h \models \mathcal{R}(C, \theta) \iff \text{dom}(h) = C \land \forall c \in C.\ h(c) = \theta
    \end{equation*}
    Equivalently: $\mathcal{R}(C, \theta) \iff \bigast_{c \in C} (c \mapsto \theta)$.

    \item \textbf{Separating Conjunction}: The heap can be split into two strictly disjoint sub-heaps, $h_1$ and $h_2$, where $h_1$ satisfies $P$ and $h_2$ satisfies $Q$. Because the Separating Conjunction is inherently defined for disjoint domains, this logic prevents spatial data races by ensuring two physical objects cannot claim the same coordinate space simultaneously.
    \begin{equation*}
        h \models P \ast Q \iff \exists h_1, h_2.\ (h_1 \perp h_2) \land (h = h_1 \uplus h_2) \land (h_1 \models P) \land (h_2 \models Q)
    \end{equation*}

    \item \textbf{Pure Logical Assertions}: Pure mathematical or set-based expressions, such as $V_{path} \setminus V_{start} \subseteq C_{stock} \cup C_{empty}$, evaluate independently of the dynamic spatial heap, relying entirely on the static geometry evaluated by the Parser.
    \begin{equation*}
        h \models \phi \iff \phi \text{ evaluates to true}
    \end{equation*}
\end{itemize}

By mapping CNC kinematics to these strict satisfaction relations, the system mathematically guarantees that if an assertion like $\mathcal{R}(V_{start}, \text{Tool}) \ast \mathcal{R}(C_{env}, \text{Environment})$ evaluates to true, the tool and environment occupy mutually exclusive, collision-free coordinate sets in $\mathbb{Z}^3$.

\subsection{Target Language and Primitives (Zero-Store Model)}

To formalize the reduction, we map physical CNC motions to the standard operations of traditional Separation Logic. Reflecting the architectural boundary where the SL Prover operates purely on pre-calculated discrete coordinates, the target language relies strictly on the following spatial primitives:

\begin{itemize}
    \item \textbf{Heap Mutation} ($[c] := \theta$): Updates the occupancy state at a specific discrete spatial address $c \in \mathbb{Z}^3$ in the spatial Heap ($h$) to a new physical object state ($\theta$).
    \item \textbf{Operational Semantics}: The mutation $[c] := \theta$ updates the spatial heap $h$ such that the discrete logical address $c$ is mapped to the object state $\theta \in \{ \text{Tool, Environment, Stock, Empty} \}$.
    \begin{equation*}
        \llbracket [c] := \theta \rrbracket = \{ (h, h') \mid h' = h[c \mapsto \theta] \} 
    \end{equation*}

    \item \textbf{Assertion} (\texttt{assert} $P$): Validates that the current heap state ($h$) satisfies the Separation Logic formula $P$.
    \item \textbf{Operational Semantics}: If $P$ is satisfied, the state remains unchanged; otherwise, the system transitions to a \textbf{fault} state, deterministically signaling a collision.
    \begin{equation*}
        \llbracket \text{assert } P \rrbracket = \{ (h, h) \mid h \models P \} \cup \{ (h, \textbf{fault}) \mid h \not\models P \}
    \end{equation*}

    \item \textbf{Finite Iteration} (\texttt{foreach} $c \in V$ \texttt{do} $\mathbb{C}$): Executes command $\mathbb{C}$ for every discrete coordinate $c$ in the finite spatial set $V$.
    \item \textbf{Operational Semantics}: This primitive models the physical progression of the tool along a path, logically represented as a sequence of discrete heap mutations over the pre-calculated swept volume $V$.
    \begin{equation*}
        \llbracket \text{foreach } c \in V \text{ do } \mathbb{C} \rrbracket = \llbracket \mathbb{C}[c_1/c]; \dots; \mathbb{C}[c_n/c] \rrbracket \text{ where } V = \{c_1, \dots, c_n\} 
    \end{equation*}
\end{itemize}

\subsection{Separation Logic Triples for G-code}

Formal verification relies on Separation Logic Triples of the form $\{P\}\mathbb{C}\{Q\}$ \cite{hoare1969axiomatic,reynolds2002separation}. Here, $P$ is the spatial precondition confirming the toolpath footprint is disjoint from environment fixtures, $\mathbb{C}$ is the discrete motion command, and $Q$ is the postcondition asserting the updated ownership of the spatial heap after the tool's displacement.

Formally, the validity of a triple is defined by the relationship between the initial heap state ($h$) and the resulting heap state ($h'$). A triple is valid if, for any state satisfying the precondition, the execution of command $\mathbb{C}$ is guaranteed not to transition to a \textbf{fault} state and results in a final state satisfying the postcondition:
\begin{equation*}
    \models \{P\}\mathbb{C}\{Q\} \iff \forall h. (h \models P \implies \neg(\llbracket \mathbb{C} \rrbracket(h) = \textbf{fault}) \land \forall h'. (h \xrightarrow{\mathbb{C}} h' \implies h' \models Q)) 
\end{equation*}
This definition ensures that safety is a formal prerequisite for state transition. A failure to satisfy the spatial disjointness required by $P$ makes the logic unsatisfiable, deterministically signaling a verification failure.

\subsection{G00 Reduction: Rapid Traversal}

Because \texttt{G00} is non-cutting, the trajectory requires the path to be strictly free of obstacles. To prevent ``ghost tool'' persistence, the reduction explicitly deallocates the tool's previous volume.

\begin{equation*}
\begin{aligned}
    & \text{G00}(V_{start}, V_{final}, V_{path}) \equiv \\
    & \mathbf{assert} \ \mathcal{R}(V_{start}, \text{Tool}) \ast \mathcal{R}(V_{path} \setminus V_{start}, \text{Empty}) ; \\
    & \text{foreach } c \in V_{path} \setminus V_{final} \text{ do } [c] := \text{Empty}; \\
    & \text{foreach } c \in V_{final} \text{ do } [c] := \text{Tool}
\end{aligned}
\end{equation*}

Operational semantics are defined strictly over the heap $h$:
\begin{align*}
    & \llbracket \text{G00}(V_{start}, V_{final}, V_{path}) \rrbracket = \mathcal{S}_{G00} \cup \mathcal{F}_{G00}\\
    & \mathcal{S}_{G00} = \left\{ (h, h') \;\middle|\; \forall c \in V_{path}.\ h(c) \notin \{\text{Environment}, \text{Stock}\} \land \forall c \in \mathbb{Z}^3.\ h'(c) = \begin{cases} \text{Tool} & \text{if } c \in V_{final} \\ \text{Empty} & \text{if } c \in V_{path} \setminus V_{final} \\ h(c) & \text{if } c \notin V_{path} \end{cases} \right\} \\
    & \mathcal{F}_{G00} = \left\{ (h, \textbf{fault}) \;\middle|\; \exists c \in V_{path}.\ h(c) \in \{\text{Environment}, \text{Stock}\} \right\} 
\end{align*}

\subsection{G01 Reduction: Linear Cutting}

This reduction involves a triple-state transformation: clearing the tool's trailing volume, transitioning traversed \texttt{Stock} to \texttt{Empty}, and allocating the new \texttt{Tool} position. 

\begin{equation*}
\begin{aligned}
    & \text{G01}(V_{start}, V_{final}, V_{path}) \equiv \\
    & \mathbf{assert} \ \mathcal{R}(V_{start}, \text{Tool}) \ast \mathcal{R}(C_{stock}, \text{Stock}) \ast \mathcal{R}(C_{empty}, \text{Empty}) \ast \mathcal{R}(C_{env}, \text{Environment}) ; \\
    & \mathbf{assert} \ V_{path} \setminus V_{start} \subseteq C_{stock} \cup C_{empty}; \\
    & \text{foreach } c \in V_{path} \setminus V_{final} \text{ do } \mathbf{if} \ [c] == \text{Stock} \ \mathbf{then} \ [c] := \text{Empty}; \\
    & \text{foreach } c \in V_{final} \text{ do } [c] := \text{Tool} 
\end{aligned}
\end{equation*}

The full operational semantics are expressed as $\llbracket \text{G01}(V_{start}, V_{final}, V_{path}) \rrbracket = \mathcal{S}_{G01} \cup \mathcal{F}_{G01}$. Successful transitions are formalized as:
\begin{equation*}
    \mathcal{S}_{G01} = \left\{ (h, h') \;\middle|\; \forall c \in V_{path}. \; h(c) \neq \text{Environment} \land \forall c \in \mathbb{Z}^3. \; h'(c) = \begin{cases} \text{Tool} & \text{if } c \in V_{final} \\ \text{Empty}  & \text{if } c \in V_{path} \setminus V_{final} \\ h(c)      & \text{if } c \notin V_{path} \end{cases} \right\}
\end{equation*}
The fault case $\mathcal{F}_{G01}$ occurs when the swept volume $V_{path}$ overlaps with non-machinable environment space:
\begin{equation*}
    \mathcal{F}_{G01} = \left\{ (h, \text{fault}) \;\middle|\; \exists c \in V_{path}. \; h(c) = \text{Environment} \right\} 
\end{equation*}

\subsection{Logical Equivalence}

The system establishes that physical CNC motion blocks are logically identical to standard loops of pointer mutations. This provides several theoretical advantages:
\begin{itemize}
    \item \textbf{Decoupling of Reasoning}: It decouples geometric calculation from logical execution. The Prover strictly dictates deterministic updates to the Spatial Heap without dynamically querying the Store.
    \item \textbf{Safety as a Gateway}: The \texttt{assert} primitive ensures physical safety is a formal prerequisite for state transition. A collision is redefined as a logical ``spatial data race'' where the separating conjunction fails to join overlapping memory domains.
    \item \textbf{Mutation as Material Consumption}: The \texttt{foreach} loops map physical machining directly to heap mutation.
\end{itemize}

\subsection{Logical Verification via Proof Rules}

Deterministic verification requires a robust proof theory. By treating space as a managed memory resource, we leverage Separation Logic to guarantee that a CNC program is collision-free.

\subsubsection{The Frame Rule}
The Frame Rule \cite{reynolds2002separation,o2004resources} allows the system to focus exclusively on the tool's immediate swept volume while protecting the workspace's spatial invariance.
\begin{equation}
    \frac{\{P\} \ \mathbb{C} \ \{Q\}}{\{P \ast R\} \ \mathbb{C} \ \{Q \ast R\}} 
\end{equation}
If a motion $\mathbb{C}$ executes safely in $P$, it will execute safely in $P \ast R$, provided the frame $R$ is mathematically disjoint from the active workspace.

\subsubsection{The Rule of Consequence}
Following the original Rule of Consequence in Hoare Logic \cite{hoare1969axiomatic}, this paper employs the rule to bridge the gap between raw geometric evaluation and logical state transitions.

\begin{equation}
    \frac{P \Rightarrow P' \quad \{P'\} \ \mathbb{C} \ \{Q'\} \quad Q' \Rightarrow Q}{\{P\} \ \mathbb{C} \ \{Q\}} 
\end{equation}
If the machine state entails a stricter geometric precondition $P'$, the logic formally transitions to the required state.

\subsection{Axiom: Spatial Disjointness and Collision}

A collision is mathematically identified when the intersection of two claimed spatial sets is non-empty.
\begin{equation}
    \frac{C_1 \cap C_2 \neq \emptyset}{\mathcal{R}(C_1, \theta_1) \ast \mathcal{R}(C_2, \theta_2) \Rightarrow \text{false}} 
\end{equation}
\begin{itemize}
    \item \textbf{Spatial Data Race}: A physical collision is redefined as a race where two distinct physical resources attempt to claim ownership of the same discrete memory address simultaneously.
    \item \textbf{Deterministic Abort}: Because the Separating Conjunction is only defined for disjoint memory addresses, an intersection makes the logic unsatisfiable, deterministically signaling a collision before physical motion occurs.
\end{itemize}

\subsection{Operational Inference Rules (Zero-Store Execution)}

Operating strictly on the pre-evaluated spatial sets emitted by the Parser, the following inference rules dictate valid state transitions within the Spatial Heap ($h$).

\subsubsection{G00 Rapid Positioning (Safe Traversal)}
Because \texttt{G00} is non-cutting, the trajectory requires the path to be strictly free of obstacles. The rule requires the entire swept volume to be \texttt{Empty} and guarantees explicit deallocation of the tool's previous volume in the post-condition.

\begin{equation}
\frac{
V_{path} \cap (C_{env} \cup C_{stock}) = \emptyset
}{
\begin{array}{c}
\{\mathcal{R}(V_{start}, \text{Tool}) \ast \mathcal{R}(V_{path} \setminus V_{start}, \text{Empty}) \ast \mathcal{R}(C_{env}, \text{Environment})\} \\
\text{G00}(V_{start}, V_{final}, V_{path}) \\
\{\mathcal{R}(V_{final}, \text{Tool}) \ast \mathcal{R}(V_{path} \setminus V_{final}, \text{Empty}) \ast \mathcal{R}(C_{env}, \text{Environment})\}
\end{array}
}
\end{equation}

\begin{itemize}
    \item \textbf{Strict Disjointness}: Following $\mathcal{S}_{G00}$ semantics, any coordinate within $V_{path}$ that overlaps with $C_{env}$ or $C_{stock}$ triggers a deterministic fault. Because the Prover evaluates this without a variable store, this rule statically ensures traversal occurs only through verified free air.
\end{itemize}

\subsubsection{G01 Linear Interpolation (Cutting and Mutation)}
Because the continuous kinematic state has been completely abstracted away by the Parser, the logic models linear cutting simply as a triple-state mutation in the Spatial Heap: clearing the tool's trailing volume, transitioning traversed \texttt{Stock} to \texttt{Empty}, and allocating the new \texttt{Tool} position. This rule accommodates trajectories passing through both workpiece material and free air.

\begin{equation}
\frac{
V_{path} \setminus V_{start} \subseteq C_{stock} \cup C_{empty}
}{
\begin{array}{c}
\{\mathcal{R}(V_{start}, \text{Tool}) \ast \mathcal{R}(C_{stock}, \text{Stock}) \ast \mathcal{R}(C_{empty}, \text{Empty}) \ast \mathcal{R}(C_{env}, \text{Environment})\} \\
\text{G01}(V_{start}, V_{final}, V_{path}) \\
\{\mathcal{R}(V_{final}, \text{Tool}) \ast \mathcal{R}(C_{stock} \setminus V_{path}, \text{Stock}) \ast \mathcal{R}(C_{empty} \cup (V_{path} \setminus V_{final}), \text{Empty}) \ast \mathcal{R}(C_{env}, \text{Environment})\}
\end{array}
}
\end{equation}

\begin{itemize}
    \item \textbf{Deallocation and Consumption}: The swept volume $V_{path} \setminus V_{final}$ is transitioned to \texttt{Empty}, representing the physical removal of \texttt{Stock} and the movement of the tool to prevent trailing ``ghost tool'' artifacts.
    \item \textbf{Path Safety}: The precondition ensures the toolpath footprint remains disjoint from the \texttt{Environment}. Any intersection with static fixtures triggers the fault transition $\mathcal{F}_{G01}$, rendering the separating conjunction unsatisfiable.
\end{itemize}

\subsection{Disjointness and Spatial Data Races}

The system formally defines a physical collision as a logical contradiction in memory ownership. Under this Zero-Store framework, physical safety is treated as a formal prerequisite for state transition rather than a post-execution outcome.

\subsubsection{Axiom: Spatial Disjointness and Collision}

A collision is mathematically identified when the intersection of two claimed spatial sets is non-empty.

\begin{equation}
\frac{C_1 \cap C_2 \neq \emptyset}{\mathcal{R}(C_1, \theta_1) \ast \mathcal{R}(C_2, \theta_2) \Rightarrow \text{false}}
\end{equation}

\begin{itemize}
    \item \textbf{Spatial Data Race:} A physical collision is redefined as a ``spatial data race'' where two distinct physical resources (e.g., \texttt{Tool} and \texttt{Environment}) attempt to claim ownership of the same discrete logical memory address simultaneously within the heap ($h$).
    \item \textbf{Deterministic Abort}: Because the Separating Conjunction ($\ast$) is only defined for disjoint memory addresses, the presence of a non-empty intersection makes the logic unsatisfiable. This transitions the system to a \texttt{false} (fault) state, deterministically signaling a collision before physical motion occurs.
\end{itemize}

\section{Case Study: Verification of Collision Scenarios}
\label{sec:case_study}

To demonstrate the deterministic verification capabilities of the proposed framework, we present a simplified case study evaluating a sequential CNC machining operation. This scenario contrasts a safe tool trajectory with an unsafe trajectory that triggers a logical spatial data race.

\subsection{Workspace Definition and Spatial Initialization}
Consider a discretized 1D workspace along the X-axis defined by the boundary $W = \{0, 1, 2, ..., 10\}$, with a scaling multiplier $\mu=1$. The physical assets are mapped to the Spatial Heap ($h$) as follows:
\begin{itemize}
    \item \textbf{Tool Geometry ($V_{tool}$):} A point tool defined by the set $\{0\}$.
    \item \textbf{Environment ($C_{env}$):} A static fixture (e.g., a clamp) located at coordinates $\{8, 9\}$.
    \item \textbf{Stock ($C_{stock}$):} The raw workpiece material occupying $\{4, 5, 6\}$.
    \item \textbf{Empty Space ($C_{empty}$):} The remaining unallocated workspace, calculated as $W \setminus (C_{env} \cup C_{stock}) = \{0, 1, 2, 3, 7, 10\}$.
\end{itemize}

The tool's initial position is at $pos = 0$. The Parser establishes the initial spatial precondition $\Sigma_{0}$ as:
$$
\Sigma_{0} \triangleq \mathcal{R}(\{0\}, Tool) * \mathcal{R}(\{4, 5, 6\}, Stock) * \mathcal{R}(\{8, 9\}, Environment) * \mathcal{R}(\{1, 2, 3, 7, 10\}, Empty)
$$

\subsection{Scenario A: Safe Traversal and Linear Cutting}
The CNC program commands the tool to safely approach the stock and perform a linear cut.
\begin{verbatim}
N10 G00 X3      (Rapid approach to safe clearance)
N20 G01 X6 F100 (Linear feed cut through the stock)
\end{verbatim}

\textbf{Evaluating N10 (G00 X3):}
The Parser statically exhausts the kinematics:
\begin{itemize}
    \item $c_{start} = 0$, $c_{final} = 3 \Rightarrow Path_{traj} = \{0, 1, 2, 3\}$
    \item $V_{path} \setminus V_{start} = \{1, 2, 3\}$
\end{itemize}
The generated SL assertion checks if the traversal path is strictly empty:
$$
\text{assert } \mathcal{R}(\{0\}, Tool) * \mathcal{R}(\{1, 2, 3\}, Empty)
$$
Because $\{1, 2, 3\} \subseteq C_{empty}$, the Prover successfully applies the G00 inference rule. The heap mutates, deallocating the trailing path to $Empty$ and allocating $\{3\}$ to $Tool$.

\textbf{Evaluating N20 (G01 X6):}
The tool cuts through the stock material. 
\begin{itemize}
    \item $c_{start} = 3$, $c_{final} = 6 \Rightarrow V_{path} = \{3, 4, 5, 6\}$
    \item $V_{path} \setminus V_{start} = \{4, 5, 6\}$
\end{itemize}
The G01 assertion requires the swept volume to consist only of $Stock$ or $Empty$ voxels:
$$
\text{assert } \{4, 5, 6\} \subseteq C_{stock} \cup C_{empty}
$$
This pure logical assertion holds true. The Prover applies the G01 rule, consuming the Stock at $\{4, 5, 6\}$ (transitioning it to Empty), and updating the Tool's location to $\{6\}$. The operation is mathematically proven safe.

\subsection{Scenario B: Deterministic Collision Detection (Spatial Data Race)}
Suppose a programming error commands the tool to rapid-traverse directly into the fixture from its current position at $X=6$.
\begin{verbatim}
N30 G00 X9      (Erroneous rapid move into the clamp)
\end{verbatim}

\textbf{Evaluating N30 (G00 X9):}
The Parser generates the discrete path:
\begin{itemize}
    \item $c_{start} = 6$, $c_{final} = 9 \Rightarrow V_{path} = \{6, 7, 8, 9\}$
\end{itemize}
According to the $\mathcal{S}_{G00}$ operational semantics, a G00 rapid positioning requires the entire trajectory to be free of obstacles. The Parser emits the following precondition for the Prover:
$$
V_{path} \cap (C_{env} \cup C_{stock}) = \emptyset
$$
However, checking this set intersection yields:
$$
\{6, 7, 8, 9\} \cap (\{8, 9\} \cup \emptyset) = \{8, 9\} \neq \emptyset
$$

Because the intersection is non-empty, the Separating Conjunction (*) fails to join the domains. The logical memory ownership is contested at coordinates $\{8, 9\}$, violating the fundamental axiom:
$$
\frac{C_{1} \cap C_{2} \neq \emptyset}{\mathcal{R}(C_{1}, \theta_{1}) * \mathcal{R}(C_{2}, \theta_{2}) \Rightarrow false}
$$
The system immediately transitions to a $fault$ state ($\mathcal{F}_{G00}$). The physical collision is deterministically detected as a Spatial Data Race prior to machine execution, effectively bypassing the computational overhead of continuous-time geometric simulations.

\section{Concurrency and Shared Workspaces}
\label{sec:concurrency}

Modern CNC environments frequently utilize multiple independent kinematic chains operating simultaneously, such as dual-spindle lathes or collaborative robotic cells. Verifying these systems requires reasoning about both independent, simultaneous motion and sequential hand-offs within shared physical workspaces. To achieve this within the architecture, the framework is extended using principles from Concurrent Separation Logic (CSL) \cite{brookes2016concurrent,o2004resources}.

\begin{figure}[h]
    \centering
    \begin{tikzpicture}[
        box/.style={
            rectangle, 
            draw, 
            align=center, 
            font=\small
        },
        local_box/.style={
            box, 
            minimum width=3.5cm, 
            minimum height=3.5cm
        },
        shared_box/.style={
            box, 
            minimum width=2.5cm, 
            minimum height=2.5cm,
            dashed 
        },
        exchange_arrow/.style={
            ultra thick, 
            <->, 
            >=stealth
        }
    ]

    \node (local1) [local_box] {
        \textbf{$h_{local\_1}$} \\
        \vspace{0.2cm}
        (Tool 1 \\ Workspace)
    };

    \node (shared) [shared_box, right=1.5cm of local1] {
        \textbf{$h_{shared}$} \\
        \vspace{0.2cm}
        Hand-off \\
        Zone
    };

    \node (local2) [local_box, right=1.5cm of shared] {
        \textbf{$h_{local\_2}$} \\
        \vspace{0.2cm}
        (Tool 2 \\ Workspace)
    };

    \draw [exchange_arrow] (shared.west) -- (local1.east);
    \draw [exchange_arrow] (shared.east) -- (local2.west);

    \node (total) [
        rectangle, 
        draw, 
        thick, 
        minimum width=13.5cm, 
        minimum height=4.7cm
    ] at (shared) {};

    \node [anchor=south west, font=\small\bfseries] at (total.north west) {TOTAL SPATIAL HEAP ($h$)};

    \end{tikzpicture}
    \caption{Heap Partitioning for Multi-Tool Execution (Concurrent Separation Logic).}
\end{figure}
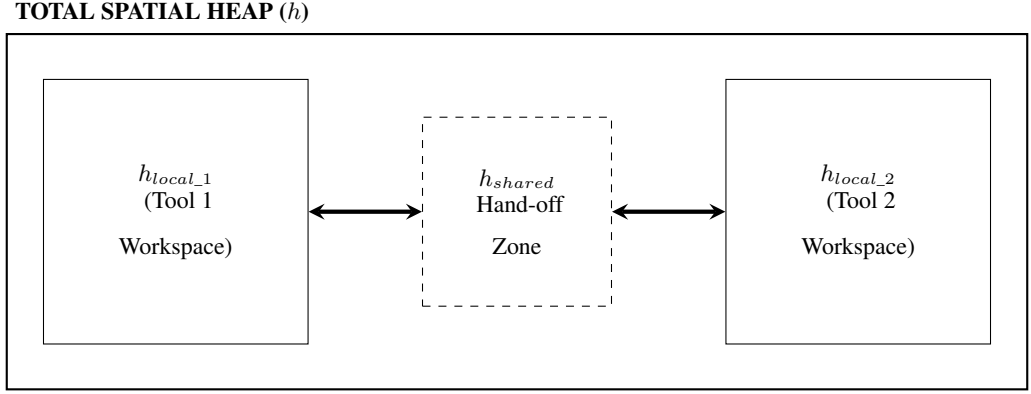

\subsection{Heap Partitioning for Multi-Tool Execution}

In a single-tool execution, the Spatial Heap ($h$) represents the entire physical workspace. To support concurrency, the heap is logically partitioned into thread-local sub-heaps representing the active space of each independent tool, alongside a shared sub-heap representing common interaction zones (e.g., a part transfer location).

Because the Separating Conjunction ($\ast$) inherently enforces disjointness, the total spatial heap is defined by the composition of these disjoint memory domains:
\begin{equation}
    h = h_{local\_1} \uplus h_{local\_2} \uplus \dots \uplus h_{shared}
\end{equation}

By partitioning the heap in this manner, explicit dynamic lock variables are unnecessary. A tool cannot physically occupy or modify a spatial address unless that address resides strictly within its respective $h_{local}$.

\subsection{Syntactic Extensions for Shared Resources}

To model sequential hand-offs where tools wait for each other to finish operating in a shared workspace, we introduce a minimal syntactic extension for Conditional Critical Regions (CCRs). 

\begin{itemize}
    \item \textbf{Resource Declaration} (\texttt{resource} $Res$ \texttt{in} $V_{shared}$): Declares a finite spatial volume $V_{shared}$ as a shared resource pool.
    \item \textbf{Resource Invariant} ($RI$): A static spatial formula that must hold true whenever the shared resource is not actively occupied by a tool thread. For instance, $RI \triangleq \mathcal{R}(V_{shared}, \text{Empty})$ asserts that the hand-off zone must be physically clear when unowned.
    \item \textbf{Critical Region} (\texttt{with} $Res$ \texttt{do} $\mathbb{C}$): A scope execution block allowing a specific G-code thread to temporarily acquire the shared resource, execute command $\mathbb{C}$, and subsequently release it.
\end{itemize}

\subsection{Formal Inference Rules for Concurrency}

Operating strictly on the pre-evaluated spatial sets, the Prover validates multi-tool interactions using the following concurrent inference rules.

\subsubsection{Disjoint Concurrency (Parallel Execution)}
When two tools operate simultaneously in entirely separate regions of the machine, their respective G-code sequences ($\mathbb{C}_1$ and $\mathbb{C}_2$) are evaluated using the Parallel Composition rule \cite{brookes2016concurrent,o2004resources}.

\begin{equation}
\frac{\{P_1\} \ \mathbb{C}_1 \ \{Q_1\} \quad \{P_2\} \ \mathbb{C}_2 \ \{Q_2\}}{\{P_1 \ast P_2\} \ \mathbb{C}_1 \parallel \mathbb{C}_2 \ \{Q_1 \ast Q_2\}}
\end{equation}

This rule guarantees that if Tool 1 and Tool 2 require disjoint spatial footprints ($P_1 \ast P_2$), they can execute safely in parallel without collision. If their swept volumes intersect in time and space, the precondition fails, deterministically detecting a multi-tool crash.

\subsubsection{Sequential Hand-offs (Critical Regions)}
For collaborative operations, a tool must temporarily claim a shared region, perform its work, and yield the space. The safe execution of this hand-off is mathematically governed by the CCR inference rule \cite{o2004resources,brookes2016concurrent}:

\begin{equation}
\frac{\{P \ast RI\} \ \mathbb{C} \ \{Q \ast RI\}}{\{P\} \ \mathbf{with} \ Res \ \mathbf{do} \ \mathbb{C} \ \{Q\}}
\end{equation}

This inference rule models the physical synchronization mechanism through memory ownership:
\begin{itemize}
    \item \textbf{Lock Acquisition via Ownership:} When Tool A enters the $\mathbf{with}$ block, it logically ``locks'' the shared physical space. Mathematically, it pulls the spatial footprint defined by $RI$ out of the shared pool and into its local heap ($P \ast RI$). Because memory ownership is strictly exclusive under the Separating Conjunction, this prevents Tool B from claiming those same spatial addresses, serving as a deterministic mutual exclusion mechanism.
    \item \textbf{Execution and Mutation:} While holding the resource, Tool A executes $\mathbb{C}$ (e.g., a \texttt{G01} motion to drop off a workpiece), temporarily violating the invariant as the space transitions from \texttt{Empty} to \texttt{Tool} or \texttt{Stock}.
    \item \textbf{Invariant Restoration and Release:} Before Tool A can safely exit the critical region and release the lock, the post-condition of its local operation must successfully restore the spatial state to satisfy the Resource Invariant ($Q \ast RI$). Physically, this requires Tool A to retract its volume from $V_{shared}$, returning the voxels back to \texttt{Empty}. Only then is the resource returned to the shared pool, allowing Tool B to subsequently acquire it.
\end{itemize}

\subsection{Scenario: Multi-Tool Concurrent Execution and Sequential Hand-off}
\label{subsec:scenario_c}

To illustrate the application of Concurrent Separation Logic (CSL) within the Zero-Store framework, we evaluate a dual-tool CNC cell. Tool A and Tool B operate in separate local workspaces but must interact within a shared hand-off zone (e.g., a part transfer station). 

\subsubsection{Heap Partitioning and Invariant Definition}
The physical workspace is discretized and logically partitioned into disjoint domains:
\begin{itemize}
    \item \textbf{Local Heaps:} $V_{localA}$ (exclusive to Tool A) and $V_{localB}$ (exclusive to Tool B).
    \item \textbf{Shared Heap:} $V_{shared}$ (the interaction zone between the two kinematics).
\end{itemize}

We declare the shared resource and its invariant ($RI$), dictating that the zone must be strictly free of any tool occupancy when not actively claimed by a thread:
$$
\mathbf{resource} \ Res_{handoff} \ \mathbf{in} \ V_{shared}
$$
$$
RI \triangleq \mathcal{R}(V_{shared}, Empty)
$$

\subsubsection{Phase 1: Disjoint Parallel Concurrency}
Both tools initially perform simultaneous cutting operations strictly within their respective local domains.
\begin{verbatim}
Thread A: G01 X10 Y10 F100 (Operating inside V_localA)
Thread B: G01 X50 Y50 F100 (Operating inside V_localB)
\end{verbatim}

The Prover evaluates this using the Parallel Composition inference rule. Let $P_A$ and $P_B$ be the local spatial preconditions for each tool:
$$
P_A \triangleq \mathcal{R}(V_{toolA}, ToolA) \ast \mathcal{R}(V_{localA} \setminus V_{toolA}, Empty)
$$
$$
P_B \triangleq \mathcal{R}(V_{toolB}, ToolB) \ast \mathcal{R}(V_{localB} \setminus V_{toolB}, Empty)
$$

Because $V_{localA} \cap V_{localB} = \emptyset$, the separating conjunction $P_A \ast P_B$ is mathematically satisfiable. The logic deterministically proves that the concurrent command $\mathbb{C}_A \parallel \mathbb{C}_B$ executes with absolute spatial safety, as the two tools are mutating perfectly disjoint segments of the global Spatial Heap.

\subsubsection{Phase 2: Sequential Hand-off via Critical Regions}
Following the parallel operations, Tool A must enter the shared zone to drop off a workpiece, requiring temporary, exclusive spatial access to $V_{shared}$.
\begin{verbatim}
Thread A:
  with Res_handoff do
    G01 X30 (Enter V_shared and execute drop-off)
    G00 X10 (Retract tool volume back into V_localA)
\end{verbatim}

\textbf{1. Lock Acquisition:} 
When Tool A enters the \texttt{with} block, it acquires the spatial lock. The Prover mathematically merges the local precondition $P_A$ with the resource invariant $RI$:
$$
P_{A}^{locked} \triangleq P_A \ast \mathcal{R}(V_{shared}, Empty)
$$
At this exact moment, if Thread B attempts to access $V_{shared}$, it will deterministically halt. The Separating Conjunction mandates exclusive memory ownership; because Thread A holds the $RI$, Thread B cannot satisfy the necessary precondition, logically preventing a multi-tool crash.

\textbf{2. Execution and Mutation:} 
Tool A executes the \texttt{G01 X30} command, moving into $V_{shared}$. This temporarily mutates the shared voxels from $Empty$ to $ToolA$, legitimately violating the invariant $RI$ while safely inside the critical region.

\textbf{3. Invariant Restoration and Release:} 
Before Tool A can exit the block, it must execute the retraction command (\texttt{G00 X10}). This physical retraction deallocates Tool A's swept volume from $V_{shared}$, transitioning the voxels back to $Empty$. This action satisfies the required post-condition $Q_A \ast RI$:
$$
\frac{\{P_A \ast RI\} \ \mathbb{C}_{enter} ; \mathbb{C}_{retract} \ \{Q_A \ast RI\}}{\{P_A\} \ \mathbf{with} \ Res_{handoff} \ \mathbf{do} \ (\mathbb{C}_{enter} ; \mathbb{C}_{retract}) \ \{Q_A\}}
$$
Upon successful verification of the retraction, the spatial lock is released. The $V_{shared}$ domain is safely returned to the shared resource pool, restoring the invariant and allowing Tool B to subsequently acquire the space.

\subsection{Extending Concurrency to Multi-Axis Kinematics}
\label{subsec:concurrency_to_multiaxis}

This formulation of Concurrent Separation Logic (CSL) naturally enables the verification of advanced multi-axis kinematics within a single machine tool.
Just as CSL safely manages multiple independent tool heads operating simultaneously within disjoint sub-heaps, the exact same theoretical foundation can be applied to a single tool and a moving workpiece.

By treating the cutting tool and the rotatable stock as two distinct, concurrent spatial ``threads'' moving through the machine workspace, 
the framework can be expanded to support simultaneous 5-axis machining without altering the underlying logic engine. This conceptual bridge between multi-tool concurrency and multi-axis dual-mutation is detailed in the following section.

\section{Multi-Axis Kinematics via Dual-Mutable Spatial Resources}
\label{sec:multi_axis_kinematics}

In a Table-Table 5-axis CNC architecture, kinematic motion is distributed between the tool spindle (linear X, Y, and Z axes) and the trunnion table (rotational B and C axes). 
To formalize this within the Spatial Heap without altering the fundamental framework of this paper, we treat multi-axis execution as the concurrent mutation of two distinct logical resources: the Tool and the Stock.

\subsection{Independent Trajectory Generation}
To maintain consistency with 3-axis logic, the SL Prover does not evaluate raw rotational or translational coordinates. 
Instead, the Parser processes the simultaneous 5-axis block (e.g., \texttt{G01 X.. Y.. Z.. B.. C..}) and splits the kinematic variables to generate two independent spatial requests prior to logical evaluation:

\begin{itemize}
    \item \textbf{Computing Linear Tool Path ($V_{path}$):} Because the tool undergoes purely translational motion, its trajectory is calculated as a standard linear swept volume. This is generated efficiently by computing the Minkowski sum of the tool's geometric profile and the linear vector of the G-code segment.
    \item \textbf{Computing Rotational Stock Path ($Stock_{path}$):} Conversely, the stock and trunnion table undergo purely rotational motion. The $Stock_{path}$ is generated by applying a rotational sweep to the current stock voxel grid around the machine's defined pivot points.
\end{itemize}

By pre-calculating these exact spatial footprints during the parsing phase, complex trigonometric kinematics are abstracted away from the verification process. 
Consequently, the Separation Logic (SL) prover receives purely discrete spatial sets, enabling formal verifications to execute within a tractable time.

\subsubsection{Kinematic Computation of the Rotational Stock Path}

To geometrically define the $Stock_{path}$ within the discrete Spatial Heap ($\mathbb{Z}^3$), the Parser isolates the rotational arguments (e.g., $a, b, c$) from the multi-axis command and applies standard rigid-body kinematic transformations. Depending on the specific Table-Table configuration of the 5-axis machine, the trunnion and rotary table utilize two of the three fundamental rotation matrices, defined as rotations about the X (A-axis), Y (B-axis), and Z (C-axis) axes:

\begin{equation}
R_A(\theta_a) = 
\begin{bmatrix}
1 & 0 & 0 \\
0 & \cos\theta_a & -\sin\theta_a \\
0 & \sin\theta_a & \cos\theta_a
\end{bmatrix}, \quad
R_B(\theta_b) = 
\begin{bmatrix}
\cos\theta_b & 0 & \sin\theta_b \\
0 & 1 & 0 \\
-\sin\theta_b & 0 & \cos\theta_b
\end{bmatrix}
\end{equation}

\begin{equation}
R_C(\theta_c) = 
\begin{bmatrix}
\cos\theta_c & -\sin\theta_c & 0 \\
\sin\theta_c & \cos\theta_c & 0 \\
0 & 0 & 1
\end{bmatrix}
\end{equation}

For a given machine configuration, let $R_{primary}$ and $R_{secondary}$ represent the active rotational axes (for instance, an A-C trunnion uses $R_A$ and $R_C$). For any coordinate point $p = (x,y,z)$ belonging to the current stock domain $Stock_{start}$, its transformed position $p'$ at specific rotational angles is given by the composite rotation matrix (assuming the machine's pivot offsets are normalized to the origin for brevity):

\begin{equation}
p'(\theta_{pri}, \theta_{sec}) = R_{primary}(\theta_{pri}) R_{secondary}(\theta_{sec}) \begin{bmatrix} x \\ y \\ z \end{bmatrix}
\end{equation}

Because a CNC command executes continuously over an interpolation interval $t \in [0,1]$, the parser directly projects this continuous rotational sweep into the discrete Spatial Heap ($\mathbb{Z}^3$). It evaluates the transformations from the starting angles to the final commanded angles, applying a boolean union to identify all discrete voxels touched by the stock during the rotation:

\begin{equation}
Stock_{path} = \{ v \in \mathbb{Z}^3 \mid \exists p \in Stock_{start}, t \in [0,1] : p'(\theta_{pri}(t), \theta_{sec}(t)) \in v \}
\end{equation}

By confining these heavy continuous trigonometric computations entirely to the Parser phase, the output $Stock_{path}$ becomes a set of discrete 3D coordinates.

\vspace{0.5cm}
\noindent \textbf{Example: Calculating a $90^\circ$ C-Axis Sweep}

To illustrate this discretization process, consider a B-C trunnion configuration where the stock contains a specific spatial voxel at coordinate $p = (10, 0, 0)$. The parser receives a multi-axis interpolation command to rotate the rotary table by 90 degrees: \texttt{G01 C90.0} (assuming the machine starts at $B=0^\circ, C=0^\circ$).

During the execution interval $t \in [0, 1]$, the B-axis remains stationary ($\theta_b(t) = 0$), meaning $R_B(0)$ is the Identity matrix. The C-axis interpolates linearly from $0$ to $\pi/2$ radians ($\theta_c(t) = t \cdot \frac{\pi}{2}$).

The continuous spatial trajectory $p'(t)$ of this specific coordinate is calculated by the parser as:
\begin{equation}
p'(t) = R_C\left(t \cdot \frac{\pi}{2}\right) \begin{bmatrix} 10 \\ 0 \\ 0 \end{bmatrix} = 
\begin{bmatrix} 
\cos(t \frac{\pi}{2}) & -\sin(t \frac{\pi}{2}) & 0 \\ 
\sin(t \frac{\pi}{2}) & \cos(t \frac{\pi}{2}) & 0 \\ 
0 & 0 & 1 
\end{bmatrix}
\begin{bmatrix} 10 \\ 0 \\ 0 \end{bmatrix}
= \begin{bmatrix} 10\cos(t \frac{\pi}{2}) \\ 10\sin(t \frac{\pi}{2}) \\ 0 \end{bmatrix}
\end{equation}

Physically, this continuous equation traces a perfect quarter-circle arc of radius 10 in the XY plane, sweeping from $(10, 0, 0)$ to $(0, 10, 0)$. 

Rather than passing this continuous trigonometric function to the Separation Logic (SL) Prover, the parser evaluates $p'(t)$ against the discrete $\mathbb{Z}^3$ spatial grid. It identifies every $1 \times 1 \times 1$ unit voxel that this arc intersects. The resulting $Stock_{path}$ subset for this specific point becomes a discrete list of integer coordinates:

\begin{equation}
Stock_{path(p)} = \{ (10,0,0), (10,1,0), (9,2,0), (9,3,0), \dots, (0,10,0) \}
\end{equation}

By expanding this operation across all voxels comprising $Stock_{start}$, the parser aggregates the total rotational swept volume. The SL Prover is completely shielded from the sine, cosine, and continuous time variables; it merely receives the final set of integer coordinates to execute its disjointness verification against the environment and the moving tool.

\subsection{Formal SL Verification of Dual Motion}
By treating both $V_{tool}$ and $C_{stock}$ as mutable domains, the SL Prover evaluates the safety of the 5-axis command through strict spatial disjointness. For a rapid, non-cutting positional move (\texttt{G00}), the trajectory requires both moving paths to be strictly free of obstacles and each other. The formal SL triple asserts this strict disjointness and guarantees the explicit deallocation of the previous volumes in the post-condition:

\[
\frac{
\begin{array}{c}
(V_{path} \cup Stock_{path}) \cap C_{env} = \emptyset \quad \text{and} \quad V_{path} \cap Stock_{path} = \emptyset
\end{array}
}
{
\begin{array}{c}
\left\{ 
\begin{array}{l}
\mathcal{R}(V_{start}, \text{Tool}) * \mathcal{R}(Stock_{start}, \text{Stock}) * \mathcal{R}(C_{env}, \text{Environment}) * \\
\mathcal{R}(\, (V_{path} \setminus V_{start}) \cup (Stock_{path} \setminus Stock_{start}) \, , \text{Empty})
\end{array}
\right\} \\
\texttt{G00}(\, V_{start}, V_{final}, V_{path}, \; Stock_{start}, Stock_{final}, Stock_{path} \,) \\
\left\{ 
\begin{array}{l}
\mathcal{R}(V_{final}, \text{Tool}) * \mathcal{R}(Stock_{final}, \text{Stock}) * \mathcal{R}(C_{env}, \text{Environment}) * \\
\mathcal{R}(\, (V_{path} \setminus V_{final}) \cup (Stock_{path} \setminus Stock_{final}) \, , \text{Empty})
\end{array}
\right\}
\end{array}
}
\]

\begin{itemize}
    \item \textbf{Strict Disjointness during Rapid Transit:} The pure logical assertions (top line) mathematically guarantee that neither the Tool nor the Stock sweeps through the static environment, and importantly, that their independent paths do not intersect each other during the rapid move. Because the SL Prover validates this prior to executing the state transition, the logic statically ensures 5-axis transit occurs strictly through verified free air.
\end{itemize}

\subsection{Formalizing Material Removal in 5-Axis}
For a linear cutting move (\texttt{G01}), material removal is logically formalized through pre-calculated spatial intersection. Because the Separating Conjunction ($*$) strictly prohibits overlapping memory claims within the dynamic heap, the Parser utilizes a pure logical assertion to evaluate absolute geometric paths before any state mutation occurs. 

The inference rule for 5-axis material removal is formally defined as follows:

\[
\frac{
\begin{array}{c}
V_{cut} = V_{path} \cap Stock_{path} \\
V_{path} \setminus V_{start} \subseteq C_{empty} \cup V_{cut} \quad \text{and} \quad Stock_{path} \setminus Stock_{start} \subseteq C_{empty} \cup V_{cut}
\end{array}
}
{
\begin{array}{c}
\{ \mathcal{R}(V_{start}, \text{Tool}) * \mathcal{R}(Stock_{start}, \text{Stock}) * \mathcal{R}(C_{empty}, \text{Empty}) * \mathcal{R}(C_{env}, \text{Environment}) \} \\
\texttt{G01}(\, V_{start}, V_{final}, V_{path}, \; Stock_{start}, Stock_{final}, Stock_{path} \,) \\
\left\{ 
\begin{array}{l}
\mathcal{R}(V_{final}, \text{Tool}) * \mathcal{R}(Stock_{final}, \text{Stock}) * \mathcal{R}(C_{env}, \text{Environment}) * \\
\mathcal{R}(\,C_{empty} \cup (V_{path} \setminus V_{final}) \cup (Stock_{path} \setminus Stock_{final}) \cup V_{cut}\, , \text{Empty}\,)
\end{array}
\right\}
\end{array}
}
\]

\vspace{0.5cm}
\noindent \textbf{The Parser-Prover Handshake:} 
The top premise isolates $V_{cut}$---the exact 3D coordinates where the tool and stock collide in absolute space---and verifies that neither moving object crashes into the environment. If the pure logical assertions hold, the SL Prover transitions the spatial heap. 

Crucially, in the post-condition, the $V_{cut}$ volume is deallocated from the Stock resource and handed over to the \texttt{Empty} spatial heap, alongside the trailing empty space abandoned by the tool ($(V_{path} \setminus V_{final})$) and the stock ($(Stock_{path} \setminus Stock_{final})$). This mechanism simulates 5-axis material consumption within space while preserving the disjointness axioms of Separation Logic.

\section{Conclusion}

The verification of CNC programs is evolving as the demands of high-speed, autonomous manufacturing outpace the scalability of repetitive geometric simulations. This paper has detailed a deterministic framework that leverages the robust mathematical foundations of Separation Logic \cite{reynolds2002separation} to provide symbolic proofs of physical trajectory safety. By conceptualizing the machine workspace as a ``Spatial Heap'' and physical collisions as logical ``Spatial Data Races,'' we have created a verification methodology that serves as a highly efficient, mathematically grounded complement to traditional testing.

The implementation of the Parser-Prover Handshake is a critical optimization that resolves continuous kinematic complexities and mechanical uncertainties into a discrete integer domain. This decoupling strategically avoids the computational intractability of quantified pointer arithmetic \cite{brotherston2018complexity}, enabling the use of local reasoning through the Frame Rule \cite{reynolds2002separation,o2004resources} to verify complex trajectories in a tractable manner \cite{cook2011tractable}. Furthermore, the adaptation of Minkowski Sum operations \cite{minarvcik2024minkowski, lien2009simple} to toolpath dilation ensures that formal safety margins are incorporated directly into the logic, accounting for real-world uncertainties like servo lag and following errors \cite{stamouli2024recursively}.

Our extensions to Concurrent Separation Logic (CSL) \cite{o2004resources,brookes2016concurrent} provide a modular framework for verifying multi-tool environments and shared robotic workspaces, treating physical hand-offs as deterministic ownership transfers. Crucially, this concurrent foundation naturally scales to advanced multi-axis kinematics. By modeling the workpiece as a dynamically mutable spatial thread, the framework natively verifies complex 5-axis operations while abstracting away heavy trigonometric inverse kinematics from the proof engine. This methodology drastically reduces the required iterative simulation cycles, offering formal safety prerequisites for state transitions in a way that mirrors mechanized semantics for additive manufacturing \cite{tekriwal2025mechanized}. 

As manufacturing systems move toward greater autonomy, the integration of such formal, manufacturer-independent logics will be essential for ensuring the reliability of the physical world. Offering a robust foundation for autonomous, less-collision manufacturing, future research will explore the potential of ``Incorrectness Separation Logic'' (ISL) \cite{raad2020local} to mathematically isolate actual collision-inducing bugs in legacy G-code, alongside the real-time integration of these provers into embedded CNC controllers for trajectory validation.

\bibliographystyle{unsrt}  
\bibliography{references}

@article{hoare1969axiomatic,
  title={An axiomatic basis for computer programming},
  author={Hoare, Charles Antony Richard},
  journal={Communications of the ACM},
  volume={12},
  number={10},
  pages={576--580},
  year={1969},
  publisher={ACM New York, NY, USA}
}

@inproceedings{reynolds2002separation,
  title={Separation logic: A logic for shared mutable data structures},
  author={Reynolds, John Charles},
  booktitle={Proceedings 17th Annual IEEE Symposium on Logic in Computer Science},
  pages={55--74},
  year={2002},
  organization={IEEE}
}

@inproceedings{brotherston2018complexity,
  title={On the complexity of pointer arithmetic in separation logic},
  author={Brotherston, James and Kanovich, Max},
  booktitle={Asian Symposium on Programming Languages and Systems},
  pages={329--349},
  year={2018},
  organization={Springer}
}

@inproceedings{cook2011tractable,
  title={Tractable reasoning in a fragment of separation logic},
  author={Cook, Byron and Haase, Christoph and Ouaknine, Jo{\"e}l and Parkinson, Matthew and Worrell, James},
  booktitle={International Conference on Concurrency Theory},
  pages={235--249},
  year={2011},
  organization={Springer}
}

@article{brookes2016concurrent,
  title={Concurrent separation logic},
  author={Brookes, Stephen and O'Hearn, Peter W},
  journal={ACM SIGLOG News},
  volume={3},
  number={3},
  pages={47--65},
  year={2016},
  publisher={ACM New York, NY, USA}
}

@inproceedings{o2004resources,
  title={Resources, concurrency and local reasoning},
  author={O’hearn, Peter W},
  booktitle={International Conference on Concurrency Theory},
  pages={49--67},
  year={2004},
  organization={Springer}
}

@inproceedings{berdine2005symbolic,
  title={Symbolic execution with separation logic},
  author={Berdine, Josh and Calcagno, Cristiano and O’hearn, Peter W},
  booktitle={Asian Symposium on Programming Languages and Systems},
  pages={52--68},
  year={2005},
  organization={Springer}
}

@inproceedings{stamouli2024recursively,
  title={Recursively feasible shrinking-horizon MPC in dynamic environments with conformal prediction guarantees},
  author={Stamouli, Charis and Lindemann, Lars and Pappas, George},
  booktitle={6th Annual Learning for Dynamics \& Control Conference},
  pages={1330--1342},
  year={2024},
  organization={PMLR}
}

@inproceedings{lien2009simple,
  title={A simple method for computing Minkowski sum boundary in 3D using collision detection},
  author={Lien, Jyh-Ming},
  booktitle={Algorithmic Foundation of Robotics VIII: Selected Contributions of the Eight International Workshop on the Algorithmic Foundations of Robotics},
  pages={401--415},
  year={2009},
  organization={Springer}
}

@inproceedings{raad2020local,
  title={Local reasoning about the presence of bugs: Incorrectness separation logic},
  author={Raad, Azalea and Berdine, Josh and Dang, Hoang-Hai and Dreyer, Derek and O’Hearn, Peter and Villard, Jules},
  booktitle={International Conference on Computer Aided Verification},
  pages={225--252},
  year={2020},
  organization={Springer}
}

@inproceedings{tekriwal2025mechanized,
  title={Mechanized Semantics for Correctness of the RS274 Additive Manufacturing Command Language},
  author={Tekriwal, Mohit and Sottile, Matthew},
  booktitle={NASA Formal Methods Symposium},
  pages={341--359},
  year={2025},
  organization={Springer}
}

@article{lozano1983spatial,
  title={Spatial planning: A configuration space approach},
  author={Lozano-Perez, Tomas and others},
  journal={IEEE Trans. Computers},
  volume={32},
  number={2},
  pages={108--120},
  year={1983}
}

@article{abrams2000computing,
  title={Computing swept volumes},
  author={Abrams, Steven and Allen, Peter K},
  journal={The Journal of Visualization and Computer Animation},
  volume={11},
  number={2},
  pages={69--82},
  year={2000},
  publisher={Wiley Online Library}
}

@inproceedings{minarvcik2024minkowski,
  title={Minkowski penalties: Robust differentiable constraint enforcement for vector graphics},
  author={Minar{\v{c}}{\'\i}k, Ji{\v{r}}{\'\i} and Estep, Sam and Ni, Wode and Crane, Keenan},
  booktitle={ACM SIGGRAPH 2024 Conference Papers},
  pages={1--12},
  year={2024}
}

@incollection{bresenham1998algorithm,
  title={Algorithm for computer control of a digital plotter},
  author={Bresenham, Jack E},
  booktitle={Seminal graphics: pioneering efforts that shaped the field},
  pages={1--6},
  year={1998},
  publisher={ACM}
}

\end{document}